\documentclass[sigplan]{acmart}
\usepackage{booktabs} % For formal tables

\usepackage{balance}       % to better equalize the last page
\usepackage{graphics}      % for EPS, load graphicx instead 
\usepackage[T1]{fontenc}   % for umlauts and other diaeresis
\usepackage{txfonts}
\usepackage{mathptmx}
\usepackage{color}
\usepackage{booktabs}
\usepackage{textcomp}

\usepackage{microtype}        % Improved Tracking and Kerning
\usepackage{ccicons}          % Cite your images correctly!
\usepackage{paralist}
\usepackage{enumitem}

\usepackage{tikz}
\usetikzlibrary{shadows,shapes}
\usepackage{pgfplots}
\pgfplotsset{compat=1.10}

\usepackage{xcolor}
\usepackage{colortbl}
\definecolor{Gray}{gray}{0.90}

\usepackage{algorithm}
\usepackage{algorithmicx}
\usepackage{algpseudocode}

\usepackage[textsize=tiny, textwidth=1.5cm]{todonotes}

\acmConference[SERIAL '17]{SERIAL 2017}
              {December 2017}{Las Vegas, Nevada USA} 
\acmYear{2017}
\copyrightyear{2017}

\begin{document}

% TODO: remove before submission, added only for todonotes
\setlength{\marginparwidth}{1.5cm}

%\allowdisplaybreaks

\title[Anonymity in Blockchain-Based Transactive Microgrids]{On the Design of Communication and Transaction Anonymity in Blockchain-Based Transactive Microgrids}

\author{Jonatan Bergquist}
\affiliation{\institution{Datarella GmbH}}
\author{Aron Laszka}
\affiliation{\institution{Vanderbilt University}}
\author{Monika Sturm}
\affiliation{\institution{Siemens Corporate Technology}}
\author{Abhishek Dubey}
\affiliation{\institution{Vanderbilt University}}

\renewcommand{\shortauthors}{J. Bergquist et al.}

%\Aron{We need to add authors (\url{https://serial17.ibr.cs.tu-bs.de/cfp.html}: ``Reviewing is single-blind.'')}

\begin{abstract}
%\textcolor{red}{The abstract needs to be rewritten}
%Transactive microgrids are emerging as the next  advancement for power grid. They provide robust capabilities for integrating distributed generation resources in communities, creat
 Transactive microgrids are emerging as a transformative solution for the problems faced by distribution system operators due to an increase in the use of distributed energy resources and a rapid acceleration in renewable energy generation, such as wind and solar power. Distributed ledgers have recently found widespread interest in this domain due to their ability to provide transactional integrity across decentralized computing nodes. However, the existing state of the art has not focused on the privacy preservation requirement of these energy systems --  the transaction level data can provide much greater insights into a prosumer's behavior compared to smart meter data. %\Aron{We could also mention safety requirements as a challenge to privacy (for stable grid control, we need to disseminate information).}\Jonatan{Resolved} 
 There are specific safety requirements in transactive microgrids to ensure the stability of the grid and to control the load. To fulfil these requirements, the distribution system operator needs transaction information from the grid, which poses a further challenge to the privacy-goals. This problem is made worse by requirement for off-blockchain communication in these networks. In this paper, we extend a recently developed trading workflow called PETra and describe our solution for communication and transactional anonymity.    
\end{abstract}

% TODO!

\maketitle

%!TEX root = paper.tex
\section{Introduction}

Transactive energy models have been proposed as a set of market based mechanisms for balancing the demand and generation of energy in communities \cite{kok2016society,cox2013structured,melton2013gridwise}.
 In this approach, customers on the same feeder (i.e. sharing a power line link) can operate in an open market, trading and exchanging generated energy locally. Distribution System Operators can be the custodian of this market, while still meeting the net demand \cite{7462854}. Blockchains have  recently emerged as a foundation for enabling the transactional service in the microgrids. For example, the Brooklyn Microgrid
(\url{brooklynmicrogrid.com}) is a peer-to-peer market for locally
generated renewable energy, which was developed by LO3 Energy as a pilot project. Similarly, RWE, and Grid Singularity have developed blockchain based solutions for incentivizing neighbors to sell excess energy to the grid and payments for electric car charging %and (c) enabling pre-paid transactions for electric bills in South Africa, respectively. 
However, those solutions do not address the requirements for off-blockchain communication network and the requirements for privacy.

 %Due to a number of challenges, however, these services have been restricted in the present to some pilot programs like Demand/Response \cite{7462854}.  On one hand, transactive energy is a decentralized power system controls problem \cite{7452738}, requiring strategic microgrid control to maintain the stability of the community and the utility. On the other hand, it is a distributed market problem where erroneous as well as malicious transactions can create a gap between demand and supply, eventually destabilizing the system.  Furthermore, this system inherently induces a distributed  infrastructure comprised of smart meters, feeders, smart inverters, utility substations, the utility central offices, and the transmission system operator, which has to provide the necessary computation fabric to support the interplay between the energy control and the fiscal market challenges.

Specifically, while blockchains provide the necessary ledger services, we still need a  communication network for sending control commands from the DSO to the prosumers as well as initiating the trade matching mechanisms.
%as described by our recent publication \textcolor{red}{\cite{Laszka17}}. 
Additionally, this communication network and the blockchain itself must preserve the privacy of the prosumers. Energy usage patterns (actual or predicted) are sensitive, personally identifiable data. Legal requirements and security considerations make it mandatory to provide a mechanism to hide the identities and transaction patterns of trade partners. Additionally, solutions must also satisfy security and safety requirements, which often conflict with privacy goals. For example, to prevent a prosumer from destabilizing the system through careless or malicious energy trading, a transactive grid must check all of the prosumer's transactions. In a decentralized system, these checks require disseminating information, which could be used to infer the prosumer's future energy consumption. 

%This paper first describes mechanisms to implement anonymity in both the communication and transactional dimensions 

In \cite{Laszka17}, we introduced {\it Privacy-preserving Energy Transactions
(PETra)}, which is our distributed-ledger based solution that
(1) enables trading energy futures in a secure and verifiable
manner, (2) preserves prosumer privacy, and (3) enables distribution system operators to regulate trading and enforce the safety rules. In this paper, we  extend the communication and transaction anonymity mechanisms. The key contributions of this paper are (a) a survey of the key concepts required for implementing the anonymity across the two dimensions, (b) a discussion on the threats that must be considered when we implement the anonymization mechanisms, and lastly (c) a discussion on implementing the anonymization extensions in PETra. 
%This paper describes privacy-aspects of the communication and transactional components of transactive microgrids. Specifically, we analyze two existing schemes for communication anonymity and two schemes for transaction anonymity. They are analyzed with respect to security and known attacks. The schemes are also judged for appropriateness in regards to their application in transactive microgrids. Based on the analysis, we propose a novel design for achieving communication and transaction anonymity in transactive microgrids.
% REDUNDANT OUTLINE
% The outline of this paper is as follows. We first summarize the transaction workflow described in \cite{Laszka17}. Thereafter, we focus on the key contributions of this paper, a discussion on the privacy challenges for both communication as well as transactions.  

% This paper introduces Privacy-preserving Energy Transactions (PETra), which is our distributed-ledger based solution that (1) enables trading energy futures in a secure and verifiable manner, (2) preserves prosumer privacy, and (3) enables DSOs to regulate trading and enforce certain safety rules. 

The outline of this paper is as follows. We first present an overview of the PETra workflow described in \cite{Laszka17} in Section \ref{sec:petra}. We then discuss the communication anonymity extensions in Section \ref{comm} and transaction anonymity in Section \ref{trans}. Section \ref{commthreat} discusses the threat vectors for the communication anonymity approach. Section \ref{transthreat} describes the transaction anonymity threats.
Finally, we provide concluding remarks in Section~\ref{sec:discussion}.

%!TEX root = paper.tex
%\vspace{-0.15in}

%\section{Analysis of State of Art}

\section{Privacy-preserving Energy Transactions}
\label{sec:petra}
%\textcolor{red}{There will be a figure here.}
\begin{figure*}[ht]
\centering
\includegraphics[width=\textwidth]{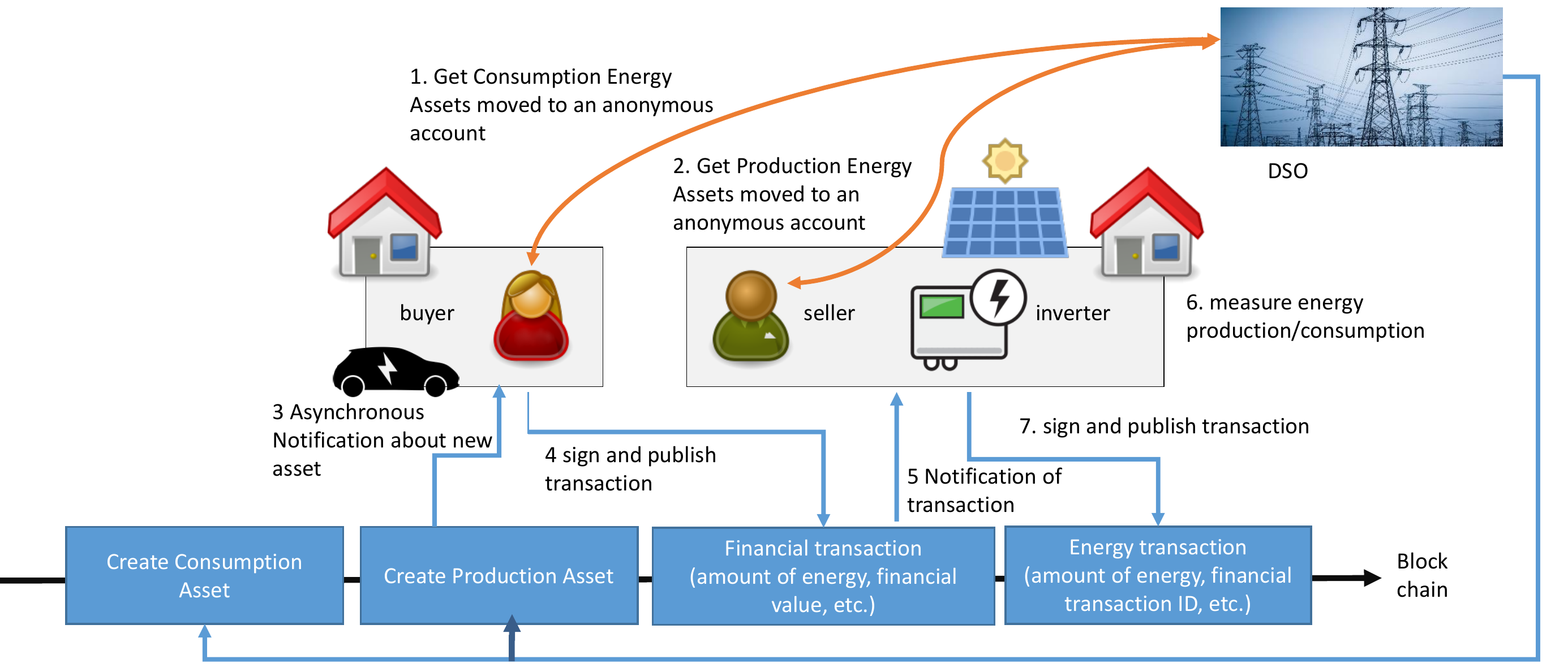}
\caption{The sequence of activities in PETra. The red arrows show off-block chain communication and blue arrows show transactions on block-chain. Producers and consumers request the DSO to allocate the energy production and consumption assets to blockchain. The consumers receive asynchronous notification about offers from producers. Thereafter, they can finalize transaction. The energy transfer happens at a later time and is also recorded in the chain. Financial transactions are also done on the blockchain. These financial transactions are later tallied with the energy transactions.}\label{fig:petrasequence}
\end{figure*}

There is a systematic pattern emerging in the domain of Internet of Things which requires transactional capabilities. Examples include transactive ride-share systems \cite{yuan2016towards}, transactive health-care systems \cite{azaria2016medrec}, and transactive energy systems described earlier in this section. As shown in Figure \ref{fig:components}, there are three separate layers of this transaction. The first layer is the distributed ledger, which is responsible for keeping track of all log of all events of interest; in the energy domain these events are trades, energy transfer and financial transactions. In case of health care domain, the events record the time of access of the health care data. The data itself is not stored in the block-chain due to the size and privacy issues. Rather, the data is stored in the second layer, which can be implemented by either a cloud or a decentralized storage service like Storj\footnote{https://storj.io/}. The third layer is the IoT layer, which is responsible for sensing and control. This third layer is typically implemented using messaging middlewares like MQTT, DDS, etc. 

\definecolor{CustomBlue}{RGB}{88, 154, 214}
\definecolor{CustomOrange}{RGB}{238, 124, 33}

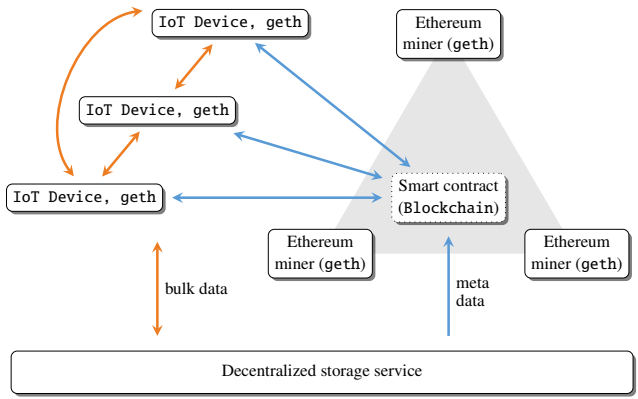
\begin{figure}
\centering
\resizebox {\columnwidth} {!} {
\centering
\begin{tikzpicture}[x=1.5cm, y=1.8cm, font=\small,
  Component/.style={fill=white, draw, align=center, rounded corners=0.1cm, drop shadow={shadow xshift=0.05cm, shadow yshift=-0.05cm, fill=black}},
  Connection/.style={<->, >=stealth, shorten <=0.15cm, shorten >=0.15cm, very thick, CustomOrange}]

\foreach \pos/\name in {0/pros1, 0.8/pros2, 1.6/pros3} {
  \node [Component] (\name) at (\pos - 4, \pos) {\texttt{IoT Device, geth}};
}

%\node [Component] (dso) at (-1, 2.6) {\texttt{IoT device, geth}};

\fill [fill=black!10] (90:1.5) -- (200:1.5) -- (340:1.5) -- (90:1.5);

\foreach \pos in {90, 200, 340} {
  \node [Component] at (\pos:1.5) {Ethereum\\miner (\texttt{geth})};
}

\node [Component, dotted] (contract) at (0, 0) {Smart contract\\(\texttt{Blockchain})};

\draw [Connection, bend left=0] (pros1) to (pros2);
\draw [Connection, bend left=0] (pros2) to (pros3);
\draw [Connection, bend left=-60] (pros3) to (pros1);

%\draw [Connection, CustomBlue, bend right=15] (dso) to (contract);
\draw [Connection, CustomBlue, bend right=0] (pros1) to (contract);
\draw [Connection, CustomBlue, bend right=0, , shorten <=0.5cm] (pros2) to (contract);
\draw [Connection, CustomBlue, bend right=0] (pros3) to (contract);

\node [Component, minimum width=10.25cm, minimum height=0.7cm] at (-1.39, -1.6) {Decentralized storage service};
\draw [Connection] (-3.2, -1.35) -- (-3.2, -0.32)  node [midway, right,black] {bulk data};
\draw [Connection, CustomBlue, ->] (0, -1.35) -- (0, -0.27)  node [midway, right, align=left, black, yshift={-0.1cm}] {meta\\[-0.2em]data};
\end{tikzpicture}
}
\vspace{-0.1in}
\caption{Components of IoT Blockchain pattern. Typically the IoT devices communicate with each other over a messaging middleware (red arrows). They also communicate with blockchain and smart contracts  (blue arrows) through clients, for example the Ethereum geth client. The miners are entities responsible for validating the events/transactions. %The distributed storage service is not shown in the figure.
}
\vspace{-0.1in}
\label{fig:components}
\end{figure}

The key aspect of this pattern is the tight integration of distributed  messaging patterns between actors and the blockchain-based communication network used for transferring transactional information. For example, in the transactive energy domain,  PETra, described in \cite{Laszka17}, involves the interactions between distribution system operator, prosumer, and a smart contract. The smart contract is  responsible for keeping track of the energy and financial assets enabling prosumers to post trade offers and exchange assets when another prosumer decides to accept.

The  algorithm of PETra uses quantised energy asset tokens\footnote{There are two kinds of energy tokens: Energy Production Asset and Energy Consumption Asset. Token attributes include power and time interval for which the token is valid.} that can represent the non-negative amount of power to be produced or consumed (for example, measured in watts),  the time interval in which energy is to be produced (or consumed) and the  last time interval in which energy is to be produced (or consumed) (Figure \ref{fig:petrasequence} describes the full sequence of activity). These assets are withdrawn and submitted to anonymized accounts on behalf of prosumers by the distribution system operator, which is also responsible for validating that the specific prosumer has the  energy capacity for feasible trades given the assets. Once the DSO posts the assets into the blockchain, prosumers can trade between themselves using these quantised assets and anonymized addresses, hiding their identity from each other. The DSO is also responsible for releasing and managing the transfer of currencies, which are represented by financial assets, which is  simply an unsigned integer value, denominated in a fiat currency. In this workflow, there are both on- and off-blockchain communications between DSO and prosumer. The off-blockchain communication is required to request the transfer of assets. On-blockchain communication occurs via filters that track the posting of assets. Similarly, prosumers also communicate which each other via blockchain to indicate when an offer has been posted and when a transaction has cleared. 

While all of the transactive IoT systems require communication and transactional anonymity there are domain-specific requirements and challenges that must be considered.  These characteristics and requirements guide us in the description of the anonymization architecture that we describe in the rest of this paper. Specifically, these characteristics are as follows: 
 (1) transactions in a microgrid must clear in bounded time and any errors must be detected\footnote{Energy trades that have an impact on real-time control (e.g., selling energy production for the near future) must be permanently recorded on the ledger \emph{in time} since grid control signals cannot be delayed.}, (2) typically, there is a dedicated communication channel available in a microgrid that connects the prosumers and the distribution system operator, (3) the set of participants in the network are fixed and known ahead of time. Thus, a discovery procedure is typically not required, and (4) even though all the transactions are anonymous there is still a need for maintaining associativity of properties like maximum generation capacity\footnote{To prevent destabilization of the grid, a producer should not be allowed to bid more than its maximum generation capacity.}, reputation scores to prosumers as they participate in trades to maximize the likelihood of success, while reducing the likelihood of jeopardizing the stability of the microgrid\footnote{A prosumer with low reputation score might have a history of not fulfilling the energy transfer obligations}. In the next two sections, we describe the mechanisms for implementing communication and transaction anonymity in this workflow.

\section{Communication and Transactive Infrastructure}

\subsection{Communication Anonymity}
\label{comm} 
The anonymous communication layer is the infrastructure upon which all
other anonymity services in PETra are built.  The goal of communication anonymity is to allow smart meters and users to exchange transactions and bids without revealing their IP-addresses or other information which can be used to identify them. In almost all cases, at the very least the Internet Service Provider (ISP) has information about the users' communications and identities. The goal of this section is to maximize the anonymity to such an extent that not even ISPs can identify users. Existing protocols for low-latency communication anonymity include onion routing ~\cite{reed1998anonymous} or the similar garlic routing \cite{Liu2014EmpiricalMA}, STAC \cite{7986314} and the decentralized Matrix protocol.\footnote{Open-federated protocol for instant messaging, Voice-over-IP and IoT communications (\url{https://matrix.org/}).} In this section, we present a brief survey of onion and garlic routing, especially with respect to application in PETra.

% \begin{figure}
%  \centering
%  \includegraphics[width=0.6\columnwidth]{onionrouting.png}
%  \caption{Conceptual visualization of onion routing between smart meters to secure communication anonymity. In the figure, smart meter A is sending a message m, with final destination B, through the onion routers.}\label{fig:onionrouting}
%  \end{figure}
 
\subsubsection{Onion and Garlic Routing}
\begin{figure}
\centering
\includegraphics[width=\columnwidth]{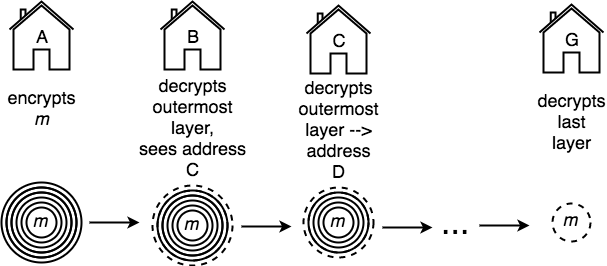}
\caption{The principle behind onion and garlic routing. The difference being that in onion routing, \textit{m} is a single message, whereas in garlic routing, \textit{m} is multiple messages packaged together.}\label{fig:garlicrouting}
\end{figure}

Onion routing is based on messages in communication being encapsulated in multiple layers of encryption and sent through a number of nodes in a network, called onion routers. It is anonymous because no single node, except for the sender and the receiver, can know the origin and the recipient of the message. In Figure \ref{fig:garlicrouting}, an example shows how smart meter A encrypts a message $m$, with final destination G, through a network of onion routers. A encrypts the message, for example a confirmation of an energy purchase, a certain number of times, along with addresses of members of the onion network. Each subsequent node, selected by the sender and specified in the different layers of encryption, decrypts one layer using its private key, revealing the next node to which the encrypted message is forwarded. Finally, the second to last node reveals the address of smart meter G and sends the still encrypted message to G, who can decrypt it safely. No single node in the network, except for the sender, knows how many times the packages is re-routed, and no node except for the sender and recipient can know their internal position in the chain of routing. Another technique for communication anonymity is called \textit{garlic routing}. It differs from onion routing in that multiple messages are encrypted together to counter tracing attacks.

In practice, the deployment of onion routing (or a variant thereof called garlic routing) in the Invisible Internet Project (I2P) works as follows. Each node in the network operates an I2P router, allowing for anonymous communications. A router is distinct from an endpoint application in that it is not a secret who runs a router. By contrast, an application is the destination for the communications and is anonymous.  This disconnect allows for a higher degree of anonymity. To communicate between routers, unidirectional tunnels are set up. The tunnels use layered encryption, meaning that each router in the tunnel only can decrypt one layer. In order to transmit a message between two routers, the sender needs to know  where to direct the message, i.e. what the address of the entry point of the receiver is. 

The I2P protocol differs from regular network communications in that, for communications to take place between routers, each router needs to know a structure called the \textit{RouterInfo}. It contains the 2048-bit ELGamal encryption key, a signing key, a certificate, timestamp, text field, signature of bundle and the contact addresses where a router can be reached. The RouterInfo is given along with something called a \textit{LeaseSet}, containing a group of tunnel entry points for a particular client destination, when the tunnel will expire, the destination itself, encryption key for end-to-end encryption of garlic messages, revocation key and a signature of the LeaseSet data. The LeaseSet identifies an application on the I2P network. The I2P protocol ensures the anonymity of its users because of the disconnect between the identities of the applications communicating over the network, and the identities of the routers. This metadata is stored in a distributed directory called the netDb, based on the Kademlia P2P-protocol, which describes a provably consistent and fault-tolerant distributed hash table. \cite{kademlia} The RouterInfo and LeaseSet data are stored on the netDb under the key derived from the SHA256 of the destination.

\subsubsection{Threat Vectors in Onion and Garlic Routing}
\label{commthreat} 
	Murdoch and Danezis \cite{1425067} show that a low-cost traffic analysis is possible of the Tor-network, theoretically and experimentally. Traffic analyses are based on tracking the forwarding of the size of a data package between computers, for example, if computer A sends a package of exactly 42 bytes to computer J, who then sends a package of exactly 42 bytes to B, it can be easily deduced that A sent a package of unknown content to computer B. This is possible because of the distribution of metadata to all routers in the Tor-network~\cite{Hopper:2010:MAN:1698750.1698753}. In what is called a timing analysis attack, an attacker tries to find a correlation between the timing of messages moving through the network to gain information about user identities and their communications. Analyses have shown that these types of attacks can be very effective over a wide range of network parameters when specific defences are not employed~\cite{Levine2004,4797313}.  To counter timing analysis attacks, the I2P network bundles multiple messages together (principle of garlic routing) and renders it more difficult to analyse~\cite{Liu2014EmpiricalMA}. Schimmer, 2009, showed that the bandwidth opportunistic peer-selection and -profiling algorithm does not prioritize anonymity in favor of performance~\cite{peerProfiling:2009}. Herrmann and Grothoff, 2011, exposed a potential weakness in anonymous HTTP-hosting done over the I2P network \cite{Herrmann2011}. The arguably only practical attack against the I2P network was done against the directory, the netDb, by Egger \textit{et al.} \cite{Egger2013}. An improvement of the protocol, aimed at Egger \textit{et al.}'s attack was suggested by Timpanaro \textit{et al.}, 2015 \cite{Timpanaro2015}. 
    
Another potential weakness of onion routing and garlic routing is that, even though the actual message is encrypted and the destinations are unknown, there is always a trace of the communication at the ISP level. The fact that a connection took place will be logged and is openly visible at the very least to the ISP. This attack can be countered in PETra by each node transacting and participating in the mixing network, regardless of the need for trading at that time. Trading of ``zero''-assets can help obfuscate the non-zero-assets of others. Another liability in onion and garlic routing can be that the legitimacy of the sender can not be immediately verified. This can be achieved by the techniques described in the section Transaction Anonymity.

\subsubsection{Proposed Solution}
Given the survey of the previous paragraphs, performing P2P energy trading in transactive grids over a garlic routing network protocol such as the I2P network provides a high amount of communication anonymity for users. Only part of the energy trading in PETra will be anonymized by garlic routing, namely the internet connections. PETra is no different from other network communications in that aspect. The particularity of the trading being local and thus IP-addresses being close, is a potential weakness that can be countered by creating ``fake'' IP-addresses. To apply garlic routing to transactive microgrids, the smart meters, prosumers, and DSO can act as onion routers, and distribution of available routers is done over netDb. In practice, this service
can be built on the free and open-source I2P software
 with private
Directory Authorities.  In this case, anonymous communication
identifiers in bids and asks correspond to public-keys that identify
I2P applications.
% ritter.vg: "run your own tor network"
% https://ritter.vg/blog-run_your_own_tor_network.html

\subsection{Transaction Anonymity}
\label{trans}
Communication anonymity is necessary but not sufficient for
anonymous trading, as the cryptographic objectives of authentication and legitimacy are not fulfilled. We suggest using cryptographic techniques from distributed ledgers, \textit{blockchains} and cryptocurrencies. The most adopted of which is the Bitcoin blockchain and currency. It allows for very simple digital cash spending but has serious privacy and anonymity flaws~\cite{Barber2012,Reid2013,apostolaki2017}. Additionally, Biryukov and Pustogarov, 2015, show that using Bitcoin over the Tor network opens an entirely new attack surface~\cite{biryukov2015}. Solutions to the tracing and identification problems identified by these researchers have been proposed and implemented in alternative cryptocurrency protocols: mixing using ring signatures and zero-knowledge proofs.~\cite{miers2013zerocoin,cryptonote} 

\subsubsection{Mixing Through Ring Signatures}
The CryptoNote protocol prevents tracing assets back to their original owners by
mixing together multiple incoming transactions and multiple outgoing
transactions. This service thus hides the connections between the
prosumers and the anonymous addresses. Mixing requires the possibility to create new wallets at will, something that is generally recommended upon any cryptocurrency transfer and it requires the existence of a sufficient number of participants in the network. These protocols enable participants to
mix assets with each other, thereby eliminating the need for a trusted
third party. Monero is an example of a cryptocurrency that provides built-in mixing services by implementing the CryptoNote protocol.~\cite{cryptoeprint:2015:1098} There are however alternative implementations of mixing protocols such as CoinShuffle~\cite{ruffing2014coinshuffle} or
Xim~\cite{bissias2014sybil}.

The CryptoNote protocol achieves two objectives:  
\begin{compactenum}
\item Untraceable transactions - \textit{for each incoming transaction all possible senders are equiprobable}. 
\item Unlinkable transactions - \textit{for any two outgoing transactions it is impossible to prove they were
sent to the same person}~\cite{cryptonote}.
\end{compactenum}

Group signatures were first introduced by Chaum and van Heyst, 1991,~\cite{Chaum1991} and then built upon by Rivest \textit{et al.}, 2001.~\cite{Rivest2001} The basis for anonymity in the CryptoNote protocol, however, is a slightly modified version of the \textit{Traceable ring signature} algorithm by Fukisaki and Suzuki, 2007.~\cite{Fujisaki2007} The algorithm allows a member of a group to send a transaction in such a way that it is impossible for a receiver to know any more about the sender than that it came from a group member without the use of a central authority.

Unlinkability is achieved by \textit{one-time ring signatures}, making use of four algorithms: \textbf{GEN, SIG, VER, LNK}. The general principle of the unconditional unlinkability is that a sender signs a transaction using a public key and a key image generated by \textbf{GEN} and produces a one-time ring signature using \textbf{SIG} and the public key pair and key image. \textbf{SIG} makes use of a non-interactive zero-knowledge proof which the verifier(s) then use to check the signature in \textbf{VER}. If the signature is valid, the verifier checks if the key image has been used in previous transactions, which mean that the same secret key was used to produce multiple signatures. She does that by running the algorithm \textbf{LNK}. Assuming that the mapping of the secret key to the key image is a one-way injection, it is certain that: \textbf{A.} The signer is not identifiable by way of recovering the secret key from the key image. \textbf{B.} The signer cannot create another key image with the same secret key. 

\begin{figure}[t]
\centering
\includegraphics[width=\columnwidth]{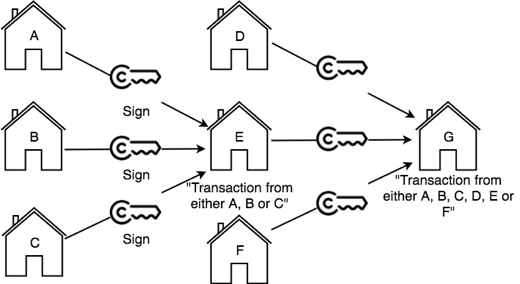}
\caption{Visualization of untraceability in ring signatures in smart meter-based energy trading and the potential deductions of origin of the transaction by a single household in the chain of signatures.}\label{fig:ringsigs}
\end{figure}

Additionally, if the receiver and sender have randomly generated, unique and new addresses, the Diffie-Hellman protocol can be used to generate a new pair of public-private keys. This is how untraceability of public keys is achieved. The sender should generate ephemeral keys for each transfer, enabling only the receiver to recover the corresponding private key. As an illustrative example, in Figure \ref{fig:ringsigs}, a schematic diagram shows households A, B and C signing a transaction since they are part of the same ring. A ring would, in reality, be many more households, not necessarily of the same microgrid. Let's assume that A is the true origin of the transaction. When E receives the transaction, the only thing that E can know with certainty is that one of A, B or C initiated the transaction. To increase the transaction anonymity further, a second, third or n rounds of ring signatures can be algorithmically imposed upon the network. With each round of signing parties, the group of potential origins grows linearly. Notably, the ring signature algorithm by Fujisaki and Suzuki, \cite{Fujisaki2007}, has been published in a peer-reviewed paper. This can be compared favorably to many cryptocurrency protocols which are simply published as white papers without any formal review-process. \cite{cryptonote}
In practice, a transaction using the mixing service should be performed in the following way to ensure anonymity:

It is also possible for household A that it paid prosumer B for energy by either disclosing the random number used in the generation of the one-time public destination key used in that transaction to B. Or she can use any other kind of zero-knowledge protocol to prove she knows the random number. The ring signatures would also allow the auditing of transactions by, for example, the DSO. This would be achieved by prosumer B giving the tracking key or truncated address to the DSO, who would then be able to link all incoming transactions to B.

\subsubsection{Mixing through Zero-Knowledge Proofs}
Zero-knowledge proofs (ZKP) are ways for a person to prove the knowledge of some specific fact to a verifier, without actually having to disclose the knowledge. Blum \textit{et al.} provided non-interactive ZKPs (NIZK) in 1988 \cite{Blum:1988:NZA:62212.62222}, where the prover and verifier don't have to interact or communicate directly with each other. The Zerocoin protocol \cite{miers2013zerocoin} outlines a way how NIZKs can achieve the untraceability objective of the previous section and it ensures that no double-spending is allowed.\footnote{Each coin in the protocol is identified uniquely by a serial number.} Zerocoin is a protocol for the decentralized mixing of coins, so that they can not be traced, or \textit{tainted}. However, senders and destinations can still be identified.~\cite{miers2013zerocoin} Luckily, Zerocash~\cite{Sasson:2014:ZDA:2650286.2650810} extends the NIZK functionality to allow for anonymous transactions, anonymous balances and coins, improved performance of transactions and sending of assets to a receivers fixed address without action required from the receiver. Zerocash makes use of a more efficient version of the NIZK, used in Zerocoin, called ZK \textit{Succinct Non-interactive ARguments of Knowledge} (zk-SNARK). 
% It features the algorithms: \textbf{Setup($1^\lambda$)}$\rightarrow \textit{params}$, \textbf{Mint(}\textit{params})$\rightarrow \textit{(c,skc)}$, \textbf{Spend}(\textit{params, c, skc, R, \textbf{C}})$\rightarrow(\pi, S)$ and \textbf{Verify}(\textit{params, $\pi$, S, R, \textbf{C}})$\rightarrow$$\{$0,1$\}$. The \textbf{Setup} function takes a secret parameter and returns a global public set of parameters and description of the set of coins. It is only executed once in the lifetime of the scheme, and the authors of the protocol recommend the usage of a trusted party for this. \textbf{Mint} is used to create coins from the coin set, given the parameters. \textbf{Spend} and \textbf{Verify} are executed by a sender and a receiver, respectively at the beginning and end of each transaction.

The Zerocash-scheme could be carried out using a simple messaging board, but would not be safe in practice since information might be manipulated or the owner of said board might collude etc. Therefore, an immutable, decentralized data storage, governed by the consensus of its peers is required to assure the secure transmission of information. The blockchain provides such a structure. 
% \begin{enumerate}[noitemsep,topsep=-\parskip]
% \item A mints a new message by generating a new random serial number $S$ of value $V$ and commits to it.
% \item The commitment, $C$, is such that only the secret message $m$ can reveal $S$.
% \item A then mints a second message by generating a new serial number $S_m$ so that the commitment $C$ is decrypted by the message $m$. 
% \item A posts the commitments to a blockchain along with the corresponding value in cryptocurrency.
% \item B has sold energy during the last timeperiod with references ($m_i,\dots m_j$) and has thereby $j-i$ number of references to scan the blockchain for.
% \item To redeem the payment, B produces a ZKP $\pi$ for the following statements:
% \begin{enumerate}[noitemsep,topsep=-\parskip]
% \item B knows a tuple ($C$) from all valid commitments available and
% \item B knows the secret $m$ such that the commitment ($C$) opens to ($S$) 
% \end{enumerate}
% \item B can now post a 'spend'-transaction containing ($S,\pi$) on the blockchain.
% \item The other users verify $\pi$ and ensure that ($S$) didn't occur already in the history of transactions.
% \item If correct, then the blockchain protocol allows B to collect cryptocurrency of value $V$. 
% \end{enumerate}
% This requires B to know all references to energy transactions that are outstanding (and coming from prosumer B) to date, and to be able to compare them to the available, valid commitments. 

\subsubsection{Threats and weaknesses in Ring Signature- and Zero-Knowledge Proof-schemes}
\label{transthreat} 
When applying either ring signatures or zero-knowledge SNARKs to PETra, potential weaknesses or attacks need to be considered.A potential threat to ring signatures is when a large amount of the unspent transactions are owned by an adversary or when insufficient amounts of signatures are included in a ring. When a prosumer A wishes to select a group of signatures to sign her transaction as well, then it is likely that she will select many of the transactions from the adversary. Assuming the adversary spends his outputs without \textit{mixing}\footnote{The number of other signatures used in the ring.}, then A's transaction is exposed as well.~\cite{monero2014} Recent research  also show that up to 65\% of Monero transactions are trivially traceable using one attack. They also exposes two more attacks that have been amended in the latest versions of the protocol, lowering the amount of transactions traceable to 20\%.~\cite{monero2014,DBLP:journals/corr/MillerMLN17} 

One of the main weaknesses of the Zerocash-based protocol is that for each private transaction, a costly zk-SNARK needs to be computed. But that is not a threat to anonymity, just a practical reason why it might be difficult to run the scheme over a congested public blockchain. In \cite{Sasson:2014:ZDA:2650286.2650810}, experiments show an average time of 3 min to create the zk-SNARK for a private transaction, verifying it takes only 8.5 ms. Another large practical drawback of Zerocash is the lack of programmability and functionality that would be required in PETra. Zhang \textit{et al.} solve some of the practical flaws and amend security issues.~\cite{zhangz} 

\subsubsection{Proposed Solution to Achieve Transaction Anonymity}
Applying the CryptoNote protocol to PETra could be done by performing both energy transactions and monetary transactions using ring signatures. They would be securely logged, tamper-proof and anonymous through the usage of a blockchain. Even though  some security flaws exists, as seen in the previous paragraph, the risk of identification, linking or tracing of transactions can be minimized by imposing a high minimum number of signatures per transaction. We also propose to connect the global transaction networks to augment the number of transactions and thereby limit the chance of deduction by elimination.

Applying ZKPs to PETra would require that a smart meter can encrypt and sign a transaction, transmit a proof of it to the blockchain and thus the receiver of the payment, without having to reveal the actual amount of energy or cost incurred to anyone but the receiver. This is achieved by the Zerocash-protocol and is implemented as a fork of the Bitcoin blockchain. Neither the receiver, nor any other participants can gain information about the transactions sent over the blockchain. To provide full functionality for PETra, the Zerocash-protocol would need to be implemented for the transmission of bids and asks as well as the already existing monetary transactions. The second implementation would need to be modified to transmit and link bids and asks to the payments ledger. A more straightforward but bloated structure would be to create transactions without monetary value to post a bid or an ask and then directly reference the final bid-/ask-transaction in the payment-transaction. 

\section{Conclusion and Discussion}
\label{sec:discussion}

Through the use of garlic routing and ring signatures, complete communication and transaction anonymity is achieved. A garlic routing network such as I2P can ensure that no usage, bid, ask or identifiable data is leaked from the system. By using ring signatures, transactions cannot be traced, but it can still be proven that a bid or an ask has been responded to and that a transaction has taken place. The design we've proposed anonymizes the whole chain of transactions, both on a network communication layer and on a distributed ledger transaction layer.

As for the DSO, it receives the same information from the smart meter
as in a non-transactive smart grid (i.e., amount of energy
produced and consumed). In
particular, since price policies are recorded on the ledger (which the
smart meters may read), each prosumer's smart meter may calculate and
send the prosumer's monthly bill to the DSO, without revealing the
prosumer's energy consumption or production. The DSO still gets aggregate information regarding load on the grid, but cannot identify individual users and their energy prosumption.

\section*{Acknowledgment}
This work was funded in part by a grant by Siemens Corporation, CT. 

% BALANCE COLUMNS
%\balance{}
%\setlength{\bibsep}{0pt plus 0ex}
%\let\oldthebibliography\thebibliography
%\let\endoldthebibliography\endthebibliography
%\renewenvironment{thebibliography}[1]{
%  \begin{oldthebibliography}{#1}
%    \setlength{\itemsep}{0.1em}
%    \setlength{\parskip}{0em}
%}
%{
%  \end{oldthebibliography}
%}

% REFERENCES FORMAT
% References must be the same font size as other body text.
\bibliographystyle{SIGCHI-Reference-Format}
\balance
\bibliography{references}

\end{document}